\documentclass[prd,preprint]{revtex4}

	\usepackage{amsmath}
	\usepackage{amsfonts}
  	\usepackage{amssymb}
	\usepackage{makeidx}
	\usepackage{amsfonts}
	\usepackage[usenames,dvipsnames]{pstricks}
	\usepackage{epsfig}
	
    \usepackage{graphicx}

	\usepackage{float}

\newcommand{\be}{\begin{equation}}
\newcommand{\ee}{\end{equation}}
\newcommand{\bea}{\begin{eqnarray}}
\newcommand{\eea}{\end{eqnarray}}

\makeindex

\begin{document}

\title{Coordinate independent approach to the calculation of the effects of local structure on the luminosity distance}
\author{Sergio Andr\'es Vallejo-Pe\~na${}^{2,3}$}
\author{Antonio Enea Romano${}^{1,2}$}

\affiliation{
${}^{1}$Theoretical Physics Department, CERN, CH-1211 Geneva 23, Switzerland\\
${}^{2}$Instituto de Fisica, Universidad de Antioquia, A.A.1226, Medellin, Colombia \\
${}^{3}${ICRANet, Piazza della Repubblica 10, I--65122 Pescara} \\
}

\begin{abstract}
Local structure can have important effects on luminosity distance observations, which could for example affect the local estimation of the Hubble constant based on low red-shift type Ia supernovae.

Using a spherically symmetric exact solution of the Eistein's equations and a more accurate expansion of the solution of the  geodesic equations, we improve the low red-shift expansion of the monopole of the luminosity distance in terms of the curvature function. Based on this we derive the coordinate independent low red-shift expansion  of the monopole of the luminosity distance in terms of  the monopole of the density contrast. The advantage of this approach is that it relates the luminosity distance directly to density observations, without any dependency on the radial coordinate choice. 

We compute the effects of different inhomogeneities on the luminosity distance, and find that the formulae in terms of the density contrast are in good agreement with  numerical calculations,  in the non linear regime are more accurate than the results obtained using linear perturbation theory, and are also more accurate than the formulae in terms of the curvature function.

\end{abstract}


\maketitle

\section{Introduction}

The luminosity distance is an observable quantity of fundamental importance for modern Cosmology, and 
it provided the first evidence of dark energy  \cite{Perlmutter:1998np,Riess:1998cb}, i.e. the late time accelerated expansion of the Universe.
Low red-shift luminosity distance observations \cite{Riess:2016jrr,Riess:2018byc,Riess:2019cxk} are also used to determine the Hubble constant $H_0$ under the assumption of spatial homogeneity, but the  value of $H_0$ obtained from local measurements is in disagreement with the  value inferred from CMB observations \cite{Riess:2016jrr,Riess:2018byc,Ade:2015xua,Aghanim:2018eyx,Riess:2019cxk}. The discrepancy has been recently claimed to be of order $4.4\sigma$ tension \cite{Riess:2019cxk}. 

The unaccounted effects of local structure on the luminosity distance could resolve this tension as shown for example in
\cite{Romano:2014iea}, and it is therefore important to study these effects. This motivates the calculation of low red-shift expansions for the luminosity distance based on using inhomogeneous exact solutions of Einstein's field equations.
The cosmological effects of inhomogeneities on different observables have been studied in \cite{Romano:2014iea,Romano:2013kua,Clarkson:2010ej,Romano:2006yc,Romano:2009xw,Ben-Dayan:2014swa,Redlich:2014gga,Romano:2009qx,EneaRomano:2011aa,Marra:2011ct,Romano:2011mx,Romano:2010nc,Krasinski:2014zza,Romano:2012yq,Romano:2012ks,Balcerzak:2013kha,Romano:2012gk,2012GReGr..44..353R,Fanizza:2013doa,RomanoChen:2014,Krasinski:2014lna,Romano:2014tya,Marra:2013rba,Romano:2015iwa,Vallejo:2017rga,Chung:2006xh}, and examples include the expansion scalar \cite{Romano:2015iwa}, number counts, \cite{Vallejo:2017rga} and the luminosity distance \cite{Romano:2010nc,Romano:2012gk,Romano:2011mx,Chung:2006xh}. 
In this paper we focus on the effects on the luminosity distance produced by the monopole of local structure modeled by a spherically symmetric solution of Einstein's equations in presence of the cosmological constant. 

A low red-shift  expansion for the monopole of the luminosity distance was  derived  in \cite{Romano:2012gk} in terms of the curvature function, and in  this paper we improve these formulae by using a more accurate expansion of the solution of the  geodesic equations \cite{Romano:2015iwa}. 
We then use these formulae to derive a new coordinate independent low red-shift expansion for the luminosity distance in terms of the monopole of the density field. 
We compare the analytic results to exact  numerical computations and perturbation theory \cite{Romano:2016utn}, finding that the formulae in terms of the density contrast are in good agreement with the numerical results and more accurate than the formulae in terms of the curvature function and the perturbative calculation. 


\section{Modeling the local Universe}
We model the monopole component of the local structure using the Lema\^itre-Tolman-Bondi (LTB) metric \cite{Lemaitre:1933,Lemaitre:1933qeLemai,Lemaitre:1931zz,Tolman:1934za,Bondi:1947av} 
\begin{eqnarray}
\label{LTBmetric} 
ds^2 = -dt^2  + \frac{R'(t,r)^2}{1 + 2\,E(r)}dr^2+R(t,r)^2
d\Omega^2 \,, 
\end{eqnarray}
where $R$ is a function of the time coordinate $t$ and the radial
coordinate $r$, $E(r)$ is an arbitrary function of $r$,
$R'(t,r)=\partial_rR(t,r)$, and we choose a system of units in which $c=8\pi G=1$. 
The Einstein's equations imply
\begin{eqnarray}
\label{Eeq1} \left({\frac{\dot{R}}{R}}\right)^2&=&\frac{2
E(r)}{R^2}+\frac{2M(r)}{R^3}+\frac{\Lambda}{3} \,,  \\
\label{Eeq2} \rho(t,r)&=&\frac{2M'}{R^2 R'} \,, 
\end{eqnarray}
where $M(r)$ is an arbitrary function of $r$ and $\dot
R(t,r)=\partial_tR(t,r)$. The luminosity distance in a LTB space-time is given by
\bea
D_L(z)&=&(1+z)^2R(t(z),r(z))\,, \label{dlz}
\eea
where $t(z)$ and $r(z)$ are the radial null geodesics, which are obtained by solving the geodesic equations \cite{Celerier:1999hp} 
\begin{eqnarray}
\frac{dr}{dz}&=&\frac{\sqrt{1+2E(r(z))}}{(1+z)\dot {R'}[t(z),r(z)]} \,,
\label{geqr} \\
\frac{dt}{dz}&=&-\,\frac{R'[t(z),r(z)]}{(1+z)\dot {R'}[t(z),r(z)]} \,.
\label{geqt} 
\end{eqnarray}

The analytical solution of eq.(\ref{Eeq1}) can be derived \cite{Zecca:2013wda,Edwards01081972} introducing a new coordinate $\eta=\eta(t,r)$, and new functions $\rho_0(r)$ and $k(r)$ given by
\bea
\frac{\partial \eta}{\partial t}|_r&=& \frac{r}{R}=\frac{1}{a}\,, \\
\rho_0(r)&=&\frac{6 M(r)}{r^3}\,, \\
k(r)&=&-\frac{2E(r)}{r^2}\,.
\eea
We will adopt, without loss of generality, the coordinate system in which $\rho_0(r)$ is a constant, the so called FLRW gauge. 
We can express eq.(2) in the form
\begin{equation}
\left(\frac{\partial a}{\partial \eta}\right)^2= -k(r)a^2 + \frac{\rho_0}{3}a + \frac{\Lambda}{3}a^4 \,.
\end{equation}

\section{Coordinate independent red-shift expansion of the luminosity distance}
Our  goal is to find an analytical formula for the luminosity distance in terms of the density contrast $\delta(z)$ at low red-shift. In order to derive the formula we take into account the metric reconstruction of the local Universe given in \cite{Vallejo:2017rga}. We follow the same procedure described in sections III and IV of \cite{Vallejo:2017rga} and we expand the curvature function $k(r)$ according to   
\begin{align}
k(r)&=k_0 + k_1 r + k_2 r^{2} + ... \,.
\end{align}

In this section we derive the formulae for the case in which $k_{0}=0$, and report the general case in appendix \ref{GF}. 

After expanding the luminosity distance in red-shift space we get
\begin{align}
D_L(z)&=D_1 z + D_2 z^2 + D_3 z^3 \,, \\
D_1&=\frac{1}{H_0}, \label{dlk1} \\ 
D_2&=\frac{4-6 \alpha  K_1 \Omega _M-3 \Omega _M}{4 H_0} \,, \label{dlk2}\\
D_3&=\frac{1}{24 H_0 \Omega _{\Lambda } \Omega _M} \Big\{-2 K_1^2 \big[2 \zeta _0+3 \Omega _{\Lambda } \Omega _M \left(4 \alpha -3 \alpha ^2 \Omega _M \left(6 \Omega _M-5\right)+3 \beta 
   \Omega _M\right)-2\big]+ \nonumber \\ & \quad {} +4 K_1 \Omega _{\Lambda } \Omega _M \big[27 \alpha  \Omega _M^2-24 \alpha  \Omega _M-2\big]+3 \Omega _{\Lambda
   } \Omega _M^2 \big[9 \Omega _M-2 \left(6 \alpha  K_2+5\right)\big]\Big\} \,, \label{dlk3}
\end{align}
where we have introduced the parameters $H_0$, $\Omega_M$, $\Omega_{\Lambda}$, $K_n$, $\alpha$, $\beta$, according to the definitions given in \cite{EneaRomano:2011aa,Romano:2015iwa,Vallejo:2017rga}. 

Note that the above formulae depend on the coordinates choice since the coefficients $D_2$ and $D_3$ given in eq.(\ref{dlk2}) and eq.(\ref{dlk3}) are expressed in terms of the coefficients $K_1$ and $K_2$, which depend on the choice of the radial coordinate $r$. It is also important to note that a similar expansion in terms of the curvature function was previously derived in \cite{Romano:2012gk}. However, the formulae we have derived is based on a more accurate expansion of the solution of the  geodesic equations \cite{Romano:2015iwa,Vallejo:2017rga}, and therefore are more precise than the previous formulae.

It is easy to check that the formulae have the correct dimensions since all the parameters are dimensionless except for $H_0$. The intermediate steps necessary to derive these expressions are rather cumbersome, for this reason the results are expressed in terms of the above mentioned parameters after making all the analytical calculations using complex simplifying routines written in Mathematica. 
This procedure  facilitates the physical interpretation of the results and ensures an immediate check of the dimensional consistency. 

As can be seen in eq.(\ref{dlk2}) the effects of the inhomogeneity start to show at second order in the red-shift expansion of the luminosity distance. Note that in absence of inhomogeneities, i.e. $K_1=K_2=0$ the obtained formulae reduce to the standard FLRW case.  

We can now use the reconstructed metric given in \cite{Vallejo:2017rga} to find the luminosity distance in terms of the density contrast. In order to do this we expand $\delta(z)$ as
\begin{equation}
\delta(z) = \delta_0 + \delta_1 z +\delta_2 z^2 \, ,
\end{equation}
and after replacing the expressions for $K_1$ and $K_2$ found in \cite{Vallejo:2017rga} into eq.(\ref{dlk2}) and eq.(\ref{dlk3}) we get
\begin{align}{\delimitershortfall=-1pt}
D_2&= -\frac{2 H_0^2 \left(3 \Omega _M-4\right) \left(3 \alpha  \Omega _M+1\right)+9 \alpha  \delta _1 \overline{H}_0^2 \overline{\Omega}_M \Omega _M}{8 H_0^3
   \left(3 \alpha  \Omega _M+1\right)}\,, \label{dldelta2} \\
D_3&= \frac{1}{640
   H_0^5 \Omega _{\Lambda } \Omega _M \left(3 \alpha  \Omega _M+1\right){}^3} \Big\{80 H_0^4 \Omega _{\Lambda } \Omega _M^2 \left(9 \Omega _M-10\right) \left(3 \alpha  \Omega _M+1\right){}^3+ \nonumber \\ & \quad {} + 3 \overline{H}_0^4 
   \overline{\Omega}_M^2 \delta _1{}^2 \big[-20 \zeta _0+1377 \alpha ^3 \Omega _{\Lambda } \Omega _M^4+30 \Omega _M^2 \left(3 \alpha ^2 \Omega
   _{\Lambda }-2 \alpha -3 \beta  \Omega _{\Lambda }\right)+  \nonumber \\ & \quad {} +324 \alpha ^2 \Omega _{\Lambda } \Omega _M^3+60 \alpha  \Omega _M+20\big]  + 8 H_0^2 \overline{H}_0^2 \overline{\Omega}_M \Omega _{\Lambda } \Omega _M\left(3 \alpha  \Omega _M+1\right) \big[\delta _1 \left(729 \alpha ^2
   \Omega _M^3+ \right. \nonumber \\ & \quad {} \left.  -72 \alpha  (10 \alpha -3) \Omega _M^2-300 \alpha  \Omega _M-20\right) -72 \alpha  \delta _2 \Omega _M \left(3 \alpha  \Omega
   _M+1\right)\big] \Big\} \,, \label{dldelta3}
\end{align}
where the parameters  $\overline{H}_0$ and $\overline{\Omega}_M$ are given in appendix \ref{AvApp}.

It is important to note that the above formulae for the luminosity distance in terms of the density contrast do not depend on the choice of radial coordinate.
It is easy to check that the formulae reduce to the FLRW luminosity distance expansion in the homogeneous limit  $\delta_1=\delta_2=0$.

\section{Testing the accuracy of the Formulae}
\begin{figure}[H]
    \includegraphics[width=.495\textwidth]{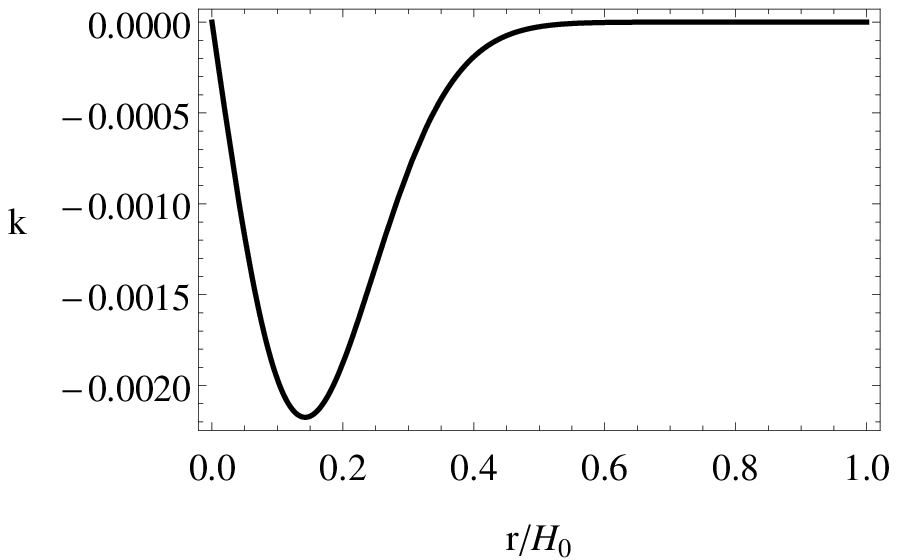}
    \includegraphics[width=.495\textwidth]{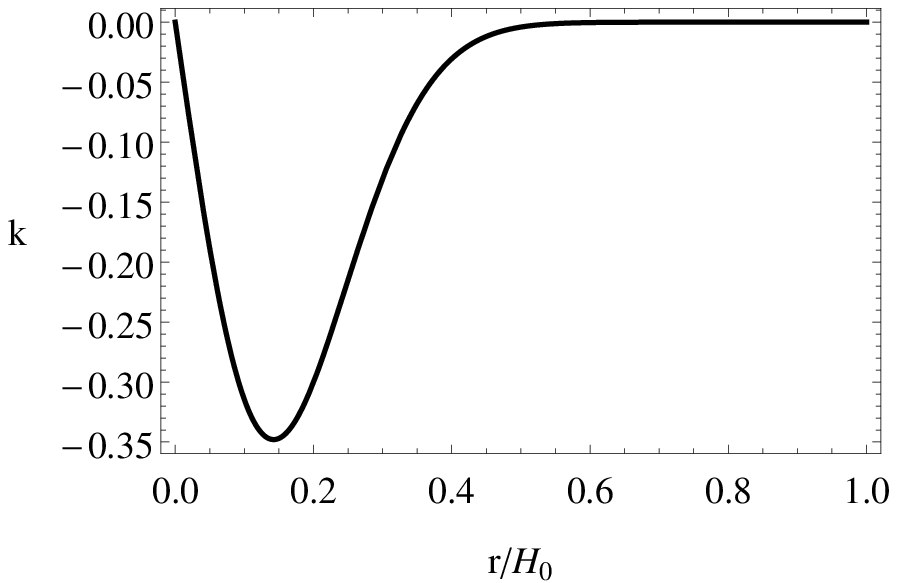}
    \caption{The function $k(r)$ defined in eq.(\ref{k}) is plotted in units of $H_0^{2}$ as a function of the radial coordinate for  $\nu= 2 \times 10^{-1}$, $\lambda=5 \times 10^{-4}$ (left) and $\nu= 2 \times 10^{-1}$, $\lambda=5 \times 10^{-2}$   (right).}
\label{knp}
\end{figure}

In order to compare our analytical formulae with numerical computations we consider inhomogeneities defined by the curvature function 
\begin{equation}
k(r) = -  \lambda H_0^2   \frac{r H_0}{\nu^2} \left(2 + \nu \, r H_0 \right) \exp \left[-\left(\frac{r H_0}{\nu} \right)^2 \right] \, , \label{k}
\end{equation}
which are shown in fig.(\ref{knp}). In order to compute the luminosity distance we first solve numerically the Einstein's eq.(\ref{Eeq1}) and the radial null geodesic equations given in eq.(\ref{geqr}) and eq.(\ref{geqt}), and then substitute the solutions of the geodesics equations in  eq.(\ref{dlz}). 
Since we are considering asymptotically flat compensated structures, the background is flat, $\overline{H}_0= H_0$ and $\overline{\Omega}_M =\Omega_M$.
  
\begin{figure}[H]
    \includegraphics[width=.495\textwidth]{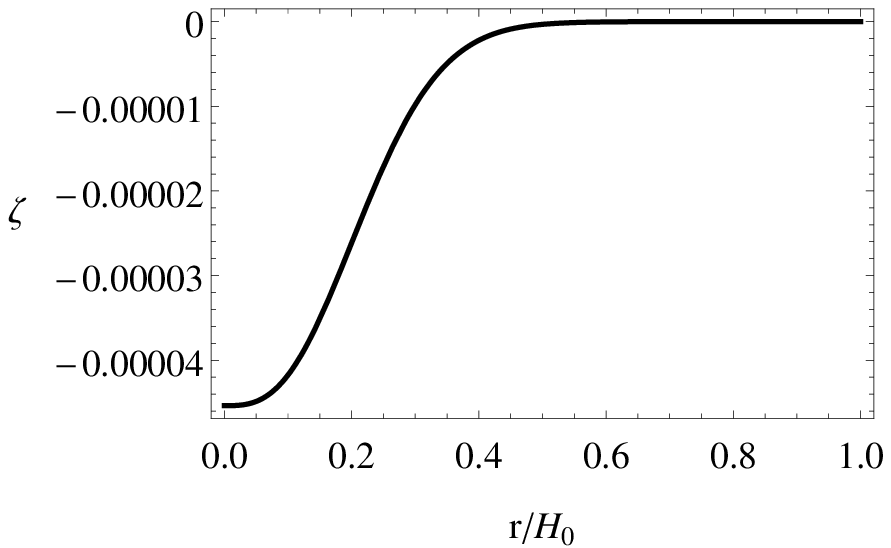}
    \includegraphics[width=.495\textwidth]{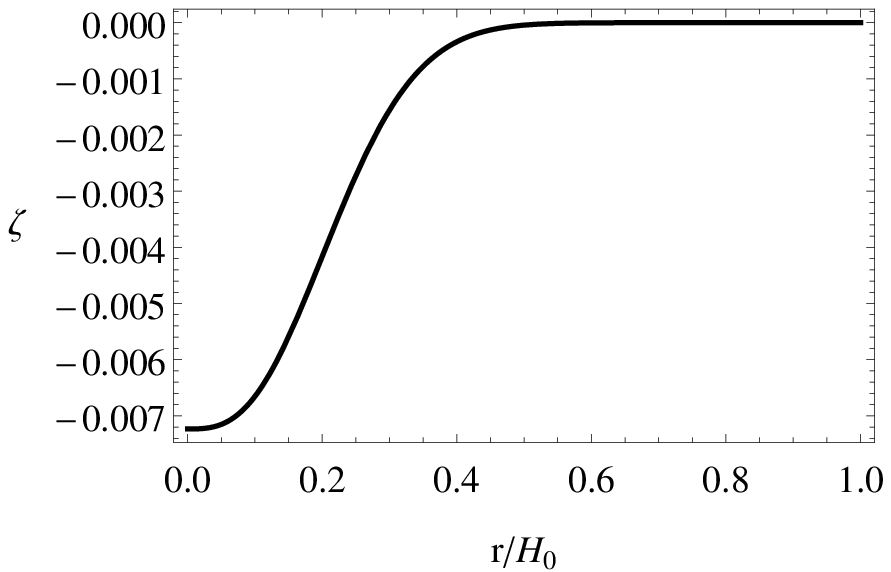}
    \caption{ The curvature perturbation $\zeta(r)$ is plotted as a function of the radial coordinate in units of $H_0^{-1}$. The left and right plots correspond to the same models shown in fig.(\ref{knp}).}   
\label{zeta}
\end{figure}

In order to check if the structure is in the linear regime we can compute the corresponding  curvature perturbation using the relation  \cite{Romano:2010nc}
\begin{equation}
    \left[ r \zeta'(r) + 1 \right]^2 = 1- r^2 k(r) \, , \label{eq:zeta}
\end{equation} 
and we plot in fig.(\ref{zeta}) this quantity for the different inhomogeneities  we consider.

\begin{figure}[H]
    \includegraphics[width=.495\textwidth]{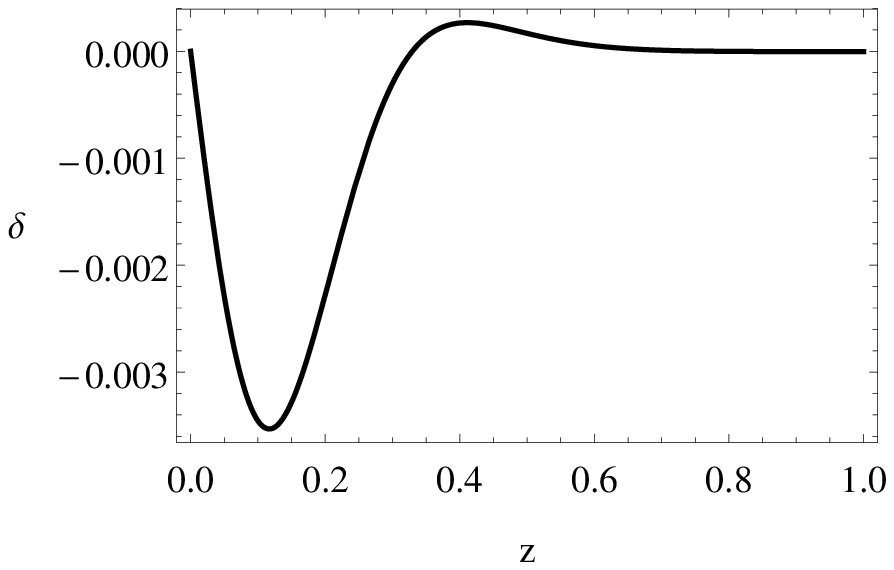}
    \includegraphics[width=.495\textwidth]{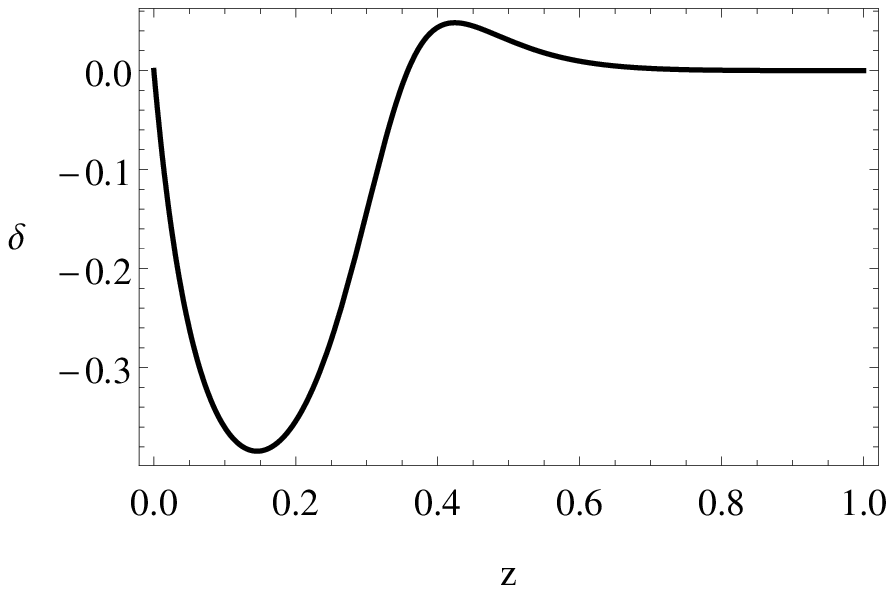}
    \caption{The density contrast $\delta(z)$ is plotted as a function of red-shift. The left and right plots correspond to the same models shown in fig.(\ref{knp}).}   
\label{delta}
\end{figure}
\begin{figure}[H]
    \centering
    \includegraphics[width=.495\textwidth]{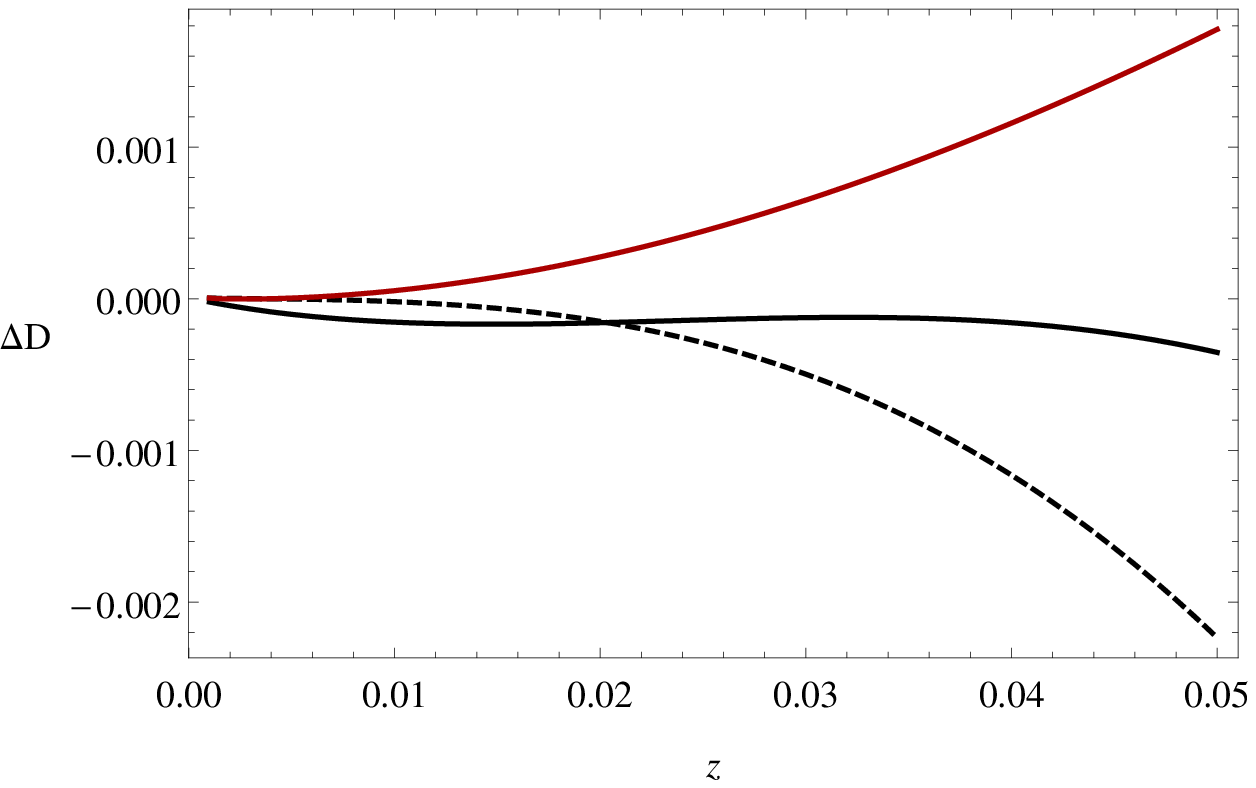}
    \includegraphics[width=.475\textwidth]{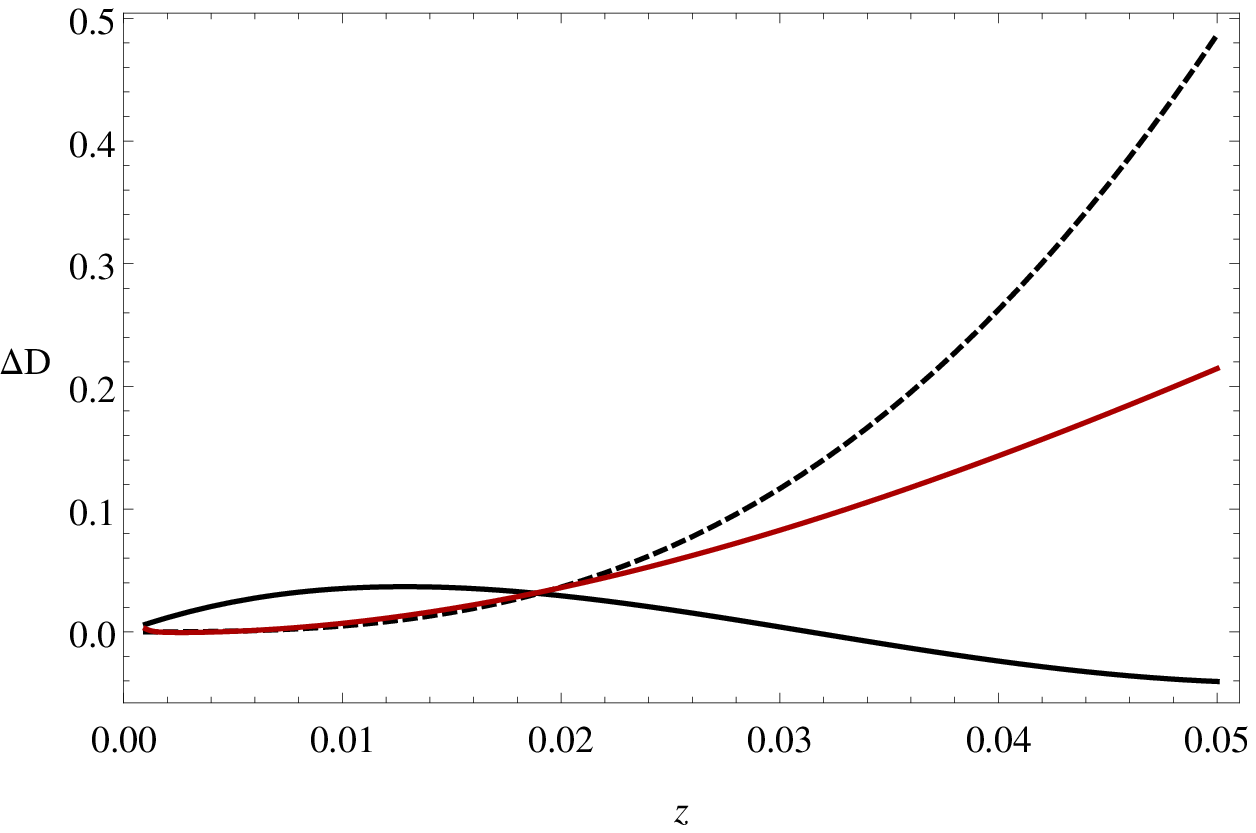}
    \caption{ The relative percentual difference $\Delta D = 100 \left(D_L^{an}/D_L^{num}-1 \right)$ is plotted as a function of red-shift. The black solid curve corresponds to the analytical formula given in terms of $\delta_1$ and $\delta_2$, the black dashed curve to the formula in terms of the curvature coefficients $K_1$ and $K_2$, and the red  curve to the perturbative approximation given in eq.(\ref{DLps}). The left and right plots correspond to the same models shown in fig.(\ref{knp}). As can be seen the coordinate independent formula is the one in best agreement with numerical results.} 
\label{DeltaDL}
\end{figure}

We also compare our formulae with the low red-shift perturbative approximation  \cite{Romano:2016utn} given by 
\begin{equation}
 D_L(z)=\overline{D}_L(z) \left[1 + \frac{1}{3} f \overline{\delta}(z)  \right] \, , \label{DLps}
\end{equation}
where 
\begin{align}
  \overline{\delta}(z)  &= \frac{3}{\chi(z)^3} \int_{0}^{z} \frac{\chi(y)^2 \delta (y)}{\overline{H}(y)} d y \, ,
\end{align}
is the volume average of the density contrast over a sphere of comoving radius $\chi(z)$,  
$\chi$, $\overline{H}$ and $\overline{D}_L$ are the comoving distance, Hubble parameter and luminosity distance of the background Universe, and $f=\overline{\Omega}_M{}^{0.55}$ is the growth rate. 
We plot in fig.(\ref{delta}) the density contrast for the different models. As can be seen from 
fig.(\ref{DeltaDL}) the formula for the luminosity distance in terms of the curvature function coefficients $K_1$ and $K_2$ is not accurate compared to the formula  in terms of the density contrast, which  is also more accurate than perturbation theory. 

\begin{figure}[H]
    \centering
    \includegraphics[width=.495\textwidth]{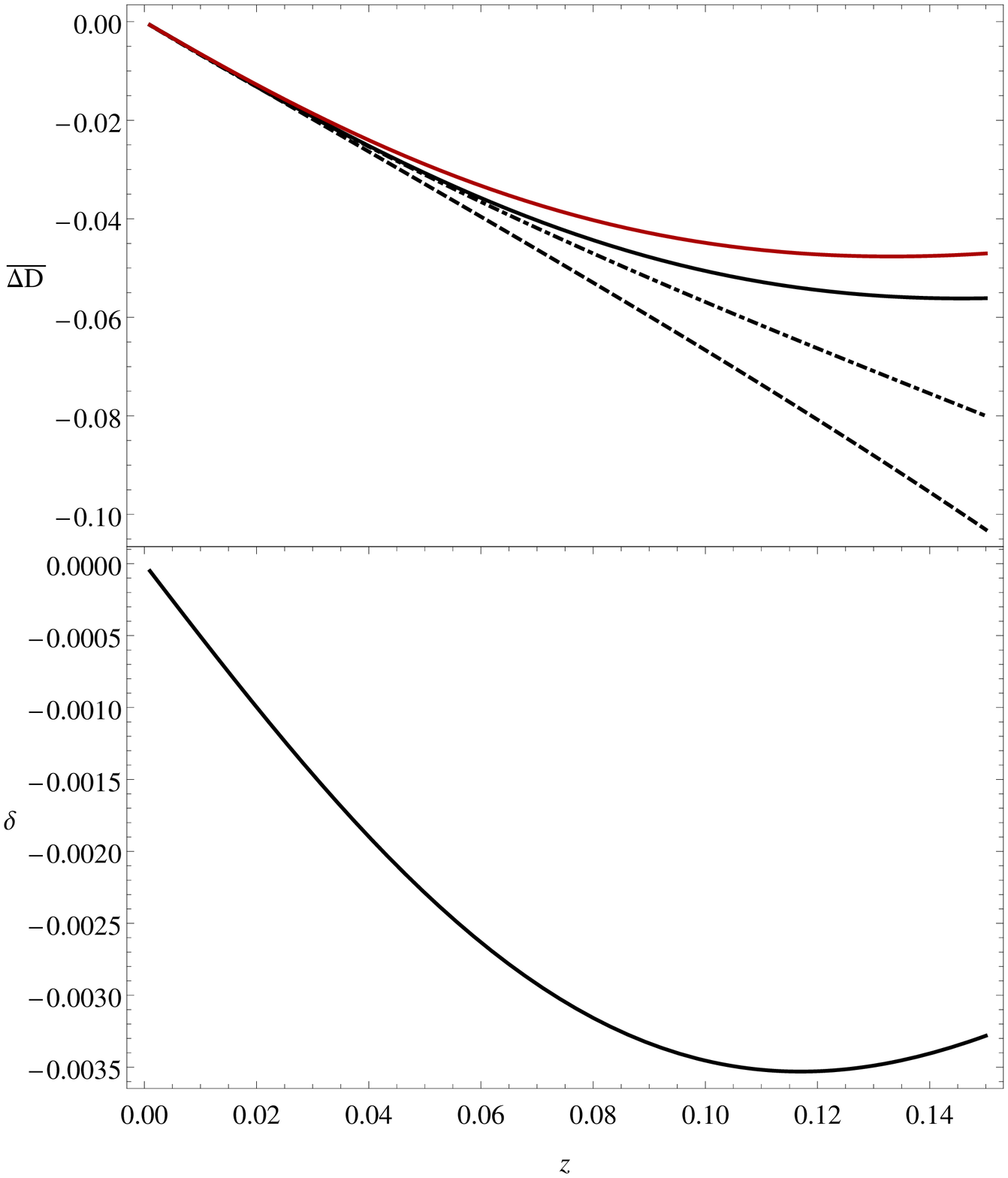}
    \includegraphics[width=.495\textwidth]{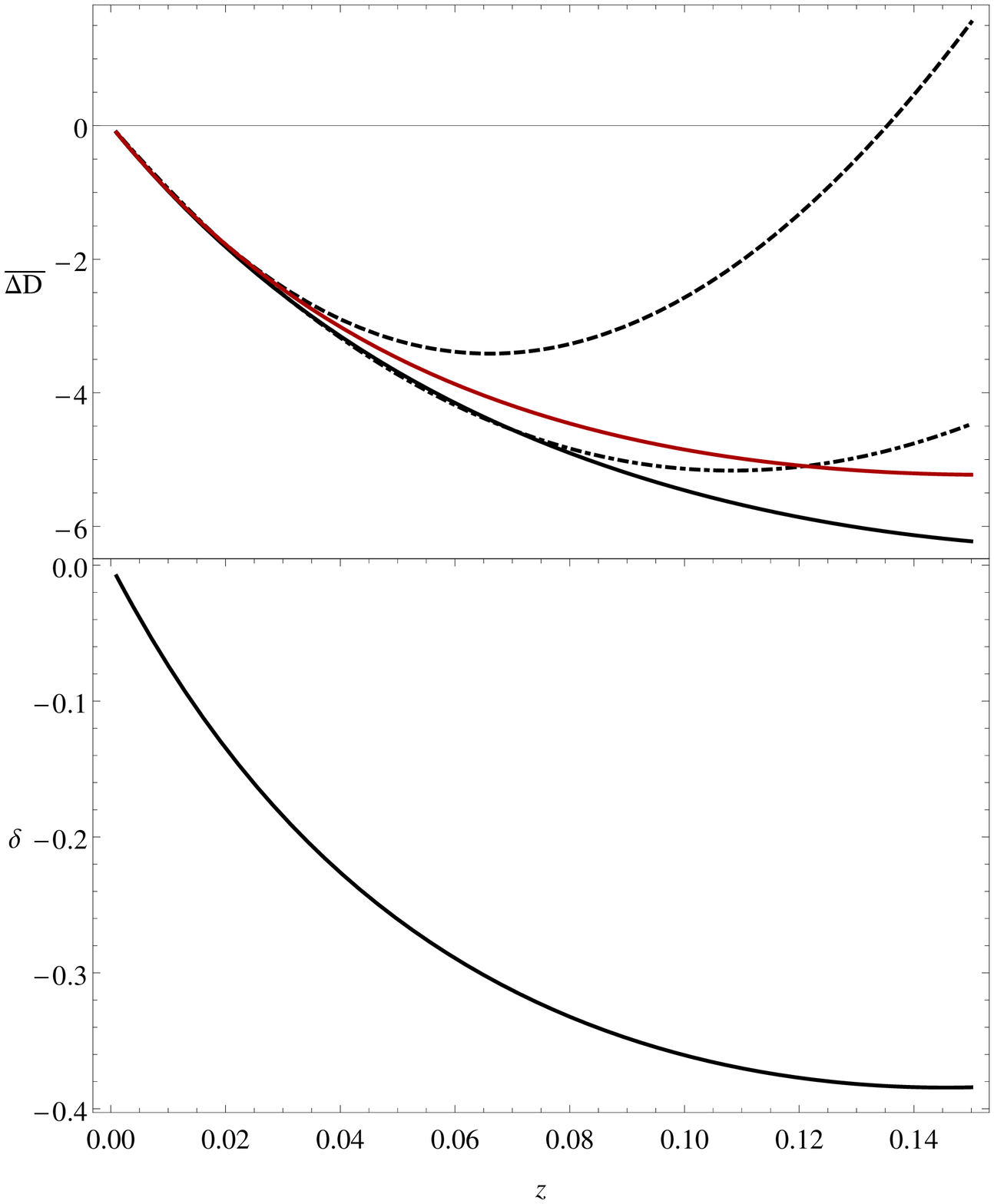}
    \caption{The relative percentual difference  $ \overline{\Delta D} = 100(D_L/\overline{D}_L-1)$ is plotted as a function of red-shift. The black solid curve corresponds to the numerical solution, the dot-dashed curve to the  coordinate independent formula in terms of the density contrast, the black dashed curve to the formula in terms of the curvature function expansion coefficients $K_1$ and $K_2$, and the red  curve to the perturbative formula given in eq.(\ref{DLps}). The left and right plots correspond to the same models shown in
    fig.(\ref{knp}). At the bottom the density contrast  is plotted as a function of red-shift for the corresponding models.} 
\label{DDbd}
\end{figure}

\section{Conclusions}

We have computed an improved low red-shift formula of the luminosity distance in terms of the curvature function using a more accurate expansion of the solution of the  geodesic equations. Based on this result we have derived for the first time a  low red-shift expansion of the luminosity distance in terms of the density contrast. The advantage of this approach is that it allows to obtain coordinate independent formulae whose coefficients are directly related to density observations, contrary to previous calculations in which the coefficients were in terms of the curvature function, and were consequently also depending on the choice of the radial coordinate.

The formulae are in good agreement with numerical calculations and are more accurate than results obtained using perturbation theory.
In the future it will be interesting to use the formulae to develop an inversion method to reconstruct the monopole of the density contrast from the monopole of the luminosity distance. The results of the inversion could be used to test the validity of the assumption of homogeneity used in the estimation of cosmological parameters, such as the Hubble constant, from local observations. 
It will also be interesting to adopt other expansion methods such as the Pad\'e approximation.
In order to consider the effects of higher multipoles of the local structure, other non spherically symmetric solutions of the Einstein's equations could be used to obtain the low red-shift expansion of the luminosity distance.

\appendix
\section{Definitions of background quantities}
\label{AvApp}

The sub-horizon volume average on constant time slices for any scalar $S(t,r)$ is defined as
\begin{align}
    \overline{S}(t) &= \frac{\int S(t,r) dV(t)}{\int dV(t)} \, ,  \label{av} \\ 
    \int dV(t) =&\int_0^{r_{h}(t)} \frac{R(t,r)^2 R'(t,r)}{\sqrt{1-k(r)r^2}} dr \, , \label{dV}
\end{align}
where $r_h(t)$ is the comoving horizon as a function of time. For compensated inhomogeneities, such as the ones we consider in this paper, the average $\overline{S}$ will is  well approximated by the asymptotic value
\begin{equation}
    \overline{S}(t)= \lim_{r\to\infty} S(t,r) \, .   \label{as}
\end{equation}
For example the background Hubble constant $\overline{H}_0$ is obtained from the volume average of the LTB Hubble parameter defined in terms of the expansion scalar 
\be
H(t,r) =  \frac{2\dot{R}(t,r)}{3R(t,r)} +  \frac{\dot{R'}(t,r)}{3R'(t,r)} \,
\ee
evaluated at present time $t_0$, $\overline{H}_0 \equiv \overline{H}(t_0)$, where $t_0=t(\eta_0,0)$.

The background parameter $\overline{\Omega}_M$ is defined in terms of the volume average of the density $\rho(t,r)$ according to
\begin{equation}
    \overline{\Omega}_M \equiv \frac{\overline{\rho}(t_0)}{3 \overline{H}_0^2} \, . 
\end{equation}

\section{General formulae}
\label{GF}
The low red-shift formulae for the luminosity distance, both the coordinate dependent and coordinate independent, are given in this appendix for the more general case in which $k_0$ is different from zero. We have used the computer algebra system provided by the Wolfram Mathematica software to derive all the formulae. 
Since $D_{1}=H_0^{-1}$ is also true for the general case we only give here the expressions for the coefficients $D_{2}$ and $D_{3}$ for simplicity, which in terms of $K_1$ and $K_2$ are given by
\allowdisplaybreaks
\begin{align}
D_2&= \frac{1}{4 H_0 \left(4 K_0^3-27 \Omega _{\Lambda } \Omega _M^2\right)} \Big\{ 8 K_0^4 \left(2 \alpha  K_1+1\right)-18 K_0 \Omega _M \big[K_1 \left(-2 \zeta _0+6 \alpha  \Omega _{\Lambda } \Omega _M+2\right)+ \nonumber \\ & \quad {} +3 \Omega _{\Lambda
   } \Omega _M\big]+27 \Omega _{\Lambda } \Omega _M^2 \big[\left(6 \alpha  K_1+3\right) \Omega _M-4\big]-4 K_0^3 \big[K_1 \left(6 \alpha  \Omega
   _M+T_0\right)+3 \Omega _M-4\big]+ \nonumber \\ & \quad {} +6 K_1 K_0^2 \left(T_0 \Omega _M-4 \zeta _0\right) \Big\} \,, \\
D_3&= \frac{1}{8 H_0 \left(4 K_0^3-27 \Omega
   _M^2 \Omega _{\Lambda }\right){}^2} \bigg\{ 64 K_0^8-64 \left(K_1 T_0+3 \Omega _M-1\right) K_0^7+16 \Big[K_1^2 T_0^2-2 K_2 T_0+ \nonumber \\& \quad {} -4 K_1 \big(6 \zeta _0+T_0 \left(1-3 \Omega
   _M\right)\big)+\Omega _M \left(9 \Omega _M-10\right)\Big] K_0^6-4 \Big[T_0 \big(-48 \zeta _0+3 T_0 \left(4 \Omega _M+ \right. \nonumber \\ & \quad {} \left. -1\right)-4\big) K_1^2+96
   \zeta _0 \left(1-3 \Omega _M\right) K_1+4 \Omega _M \big(T_0 \left(9 \Omega _M-8\right)+8\big) K_1+12 \big(18 \Omega _{\Lambda } \Omega _M^2+ \nonumber \\ & \quad {} +K_2
   \left(4 \zeta _0-T_0 \Omega _M\right)\big)\Big] K_0^5+12 \Big[\big(48 \zeta _0^2+4 \left(T_0 \left(3-12 \Omega _M\right)+4\right) \zeta _0+ \nonumber \\ & \quad {} +T_0
   \Omega _M \left(T_0 \left(3 \Omega _M-2\right)+11\right)\big) K_1^2+4 \Omega _M \big(-2 \zeta _0 \left(9 \Omega _M-8\right)-8 \Omega _{\Lambda
   }+\Omega _M \left(9 T_0 \Omega _{\Lambda }+ \right. \nonumber \\ & \quad {} \left. +6\right)-4\big) K_1+24 \Omega _M \big(K_2 \left(\zeta _0-1\right)+3 \Omega _M \left(3 \Omega _M-1\right)
   \Omega _{\Lambda }\big)\Big] K_0^4+36 \Big[\big(\left(12+ \right. \nonumber \\ & \quad {} \left. -48 \Omega _M\right) \zeta _0^2+2 \Omega _M \left(T_0 \left(6 \Omega _M-4\right)+9\right)
   \zeta _0+\Omega _M \left(T_0 \left(-7 \Omega _M+3 \Omega _{\Lambda }+3\right)+12\right)\big) K_1^2+ \nonumber \\ & \quad {} +4 \big(18 \zeta _0+T_0 \left(3-9 \Omega
   _M\right)+2\big) \Omega _M^2 \Omega _{\Lambda } K_1+6 \Omega _M^2 \big(K_2 T_0+\left(10-9 \Omega _M\right) \Omega _M\big) \Omega _{\Lambda
   }\Big] K_0^3+ \nonumber \\ & \quad {} +108 \Omega _M \Big[\big(4 \left(3 \Omega _M-2\right) \zeta _0^2+\left(-14 \Omega _M+6 \Omega _{\Lambda }+6\right) \zeta _0+\Omega _M
   \left(T_0 \Omega _{\Lambda }+4\right)\big) K_1^2+ \nonumber \\ & \quad {} +\Omega _M \big(\zeta _0 \left(24-72 \Omega _M\right)+\Omega _M \left(T_0 \left(9 \Omega
   _M-8\right)+8\right)\big) \Omega _{\Lambda } K_1+3 \Omega _M \Omega _{\Lambda } \big(9 \Omega _{\Lambda } \Omega _M^2+ \nonumber \\ & \quad {} +K_2 \left(4 \zeta _0-T_0
   \Omega _M\right)\big)\Big] K_0^2+2 \Big[\alpha ^2 \big(18 \Omega _M^2-3 \left(8 K_0+5\right) \Omega _M+8 K_0 \left(K_0+1\right)\big)
   K_1^2+ \nonumber \\ & \quad {} +\left(\beta  K_1^2+2 \alpha  K_2\right) \left(2 K_0-3 \Omega _M\right)\Big] \left(4 K_0^3-27 \Omega _M^2 \Omega _{\Lambda }\right){}^2-324 \Omega _M^2 \Omega _{\Lambda } \Big[\left(T_0 \Omega _M-3\right) K_1^2+ \nonumber \\ & \quad {} -2 \Omega _M \big(-3 \Omega _M+\zeta
   _0 \left(9 \Omega _M-8\right)+4 \Omega _{\Lambda }+2\big) K_1+3 \Omega _M \big(2 K_2 \left(\zeta _0-1\right)+3 \Omega _M \left(3 \Omega _M+ \right. \nonumber \\ & \quad {} \left. -1\right)
   \Omega _{\Lambda }\big)\Big] K_0 +243
   \Omega _M^3 \Omega _{\Lambda } \Big[-4 \left(\zeta _0-1\right) K_1^2-8 \Omega _M \Omega _{\Lambda } K_1+3 \Omega _M^2 \left(9 \Omega _M-10\right)
   \Omega _{\Lambda }\Big]+ \nonumber \\ & \quad {} +4 \alpha  K_1 \left(4 K_0^3-27 \Omega _M^2 \Omega _{\Lambda }\right) \Big[16 K_0^5-8 \left(K_1 T_0+6 \Omega _M-2\right)
   K_0^4+4 \big(\Omega _M \left(9 \Omega _M-8\right)+ \nonumber \\ & \quad {} +K_1 \left(-12 \zeta _0+T_0 \left(6 \Omega _M-2\right)-1\right)\big) K_0^3-3 \big(36 \Omega
   _{\Lambda } \Omega _M^2+K_1 \{\zeta _0 \left(16-48 \Omega _M\right)+ \nonumber \\ & \quad {} +\Omega _M [T_0 \left(6 \Omega _M-5\right)+4]\}\big) K_0^2-18
   \Omega _M \big(6 \left(1-3 \Omega _M\right) \Omega _M \Omega _{\Lambda }+K_1 \left(-2 \Omega _M+ \right. \nonumber \\ & \quad {} \left. +\zeta _0 \left(6 \Omega _M-5\right)+3 \Omega _{\Lambda
   }+1\right)\big) K_0+27 \Omega _M^2 \left(2 K_1+\left(8-9 \Omega _M\right) \Omega _M\right) \Omega _{\Lambda }\Big] \bigg\} \,.  
\end{align}

The coordinate independent coefficients $D_{2}$ and $D_{3}$ are given by
\begin{align}
D_2&= \frac{1}{8 H_0^3 \Omega _M \mathcal{A} } \bigg\{ \delta _1 \overline{H}_0^2 \overline{\Omega}_M \Big[-8 \alpha  K_0^4+18 K_0 \Omega _M \left(-\zeta _0+3 \alpha  \Omega _{\Lambda } \Omega _M+1\right)+2 K_0^3
   \left(6 \alpha  \Omega _M+ \right. \nonumber \\ & \quad {} \left. +T_0\right)+3 K_0^2 \left(4 \zeta _0-T_0 \Omega _M\right)-81 \alpha  \Omega _{\Lambda } \Omega _M^3\Big]+2 H_0^2 \Omega _M
   \left(2 K_0-3 \Omega _M+4\right) \mathcal{A} \bigg\}\,, \\
D_3&= -\frac{1}{640 H_0^5 \Omega _M^2 \mathcal{A}{}^3} \Bigg\{ \overline{\Omega}_M^2 \delta _1^2 \bigg[4352 \alpha ^3 K_0^{11}-64 \alpha ^2 \big(51 T_0+4 \left(51 \Omega _M \alpha +5 \alpha +6\right)\big)
   K_0^{10}+ \nonumber \\ & \quad {} +16 \Big(\alpha  \big\{51 T_0^2+12 \left(51 \Omega _M \alpha +5 \alpha +4\right) T_0+4 \alpha  [-306 \zeta _0+9 \Omega _M
   \left(17 \alpha  \Omega _M+16\right)+  \nonumber \\ & \quad {}  +40]\big\}-40 \beta \Big) K_0^9-4 \Big(17 T_0^3+12 \left(51 \Omega _M \alpha +5 \alpha
   +2\right) T_0^2+4 \alpha  \big\{-612 \zeta _0+ \nonumber \\ & \quad {} +9 \Omega _M \left(51 \alpha  \Omega _M+32\right)+80\big\} T_0-720 \beta  \Omega _M+16 \alpha 
   \big\{-18 \zeta _0 \left(51 \Omega _M \alpha +5 \alpha +4\right)+ \nonumber \\ & \quad {} +9 \alpha  \Omega _M [30 \Omega _M+3 \left(51 \alpha  \Omega
   _M+4\right) \Omega _{\Lambda }+34]+25\big\}\Big) K_0^8+4 \Big(\left(51 \Omega _M+5\right) T_0^3+ \nonumber \\ & \quad {} +\big\{-306 \zeta _0+9 \Omega _M
   \left(51 \alpha  \Omega _M+16\right)+50\big\} T_0^2+4 \big\{-36 \zeta _0 \left(51 \Omega _M \alpha +5 \alpha +2\right)+ \nonumber \\ & \quad {} +9 \alpha  \Omega _M
   [60 \Omega _M+6 \left(51 \alpha  \Omega _M+4\right) \Omega _{\Lambda }+73]+25\big\} T_0-720 \beta  \Omega _M \left(\Omega
   _M+\Omega _{\Lambda }-1\right)+ \nonumber \\ & \quad {} +24 \alpha  \big\{54 \alpha  \left(51 \alpha  \Omega _{\Lambda }+2\right) \Omega _M^3+9 \alpha  [-51
   \zeta _0+6 (5 \alpha +8) \Omega _{\Lambda }+29] \Omega _M^2+\left(90 \alpha -288 \zeta _0+ \right. \nonumber \\ & \quad {} \left. +137\right) \Omega _M+2 \zeta _0 \left(153
   \zeta _0-40\right)\big\}\Big) K_0^7-3 \Big(66096 \alpha ^3 \Omega _{\Lambda } \Omega _M^4+864 \alpha  \big\{54 \alpha  \Omega _{\Lambda
   }+ \nonumber \\ & \quad {} +T_0 \left(51 \alpha  \Omega _{\Lambda }+2\right)\big\} \Omega _M^3+3 \big\{17 T_0^3+12 \left(51 \alpha  \Omega _{\Lambda }+10\right)
   T_0^2+16 \alpha  [-153 \zeta _0+18 (5 \alpha + \nonumber \\ & \quad {} +6) \Omega _{\Lambda }+92] T_0+3296 \alpha -288 [5 (\beta -3 \alpha
   ^2) \Omega _{\Lambda }+2 \alpha  \zeta _0 \left(51 \alpha  \Omega _{\Lambda }+10\right)]\big\} \Omega _M^2+ \nonumber \\ & \quad {} +8 \big\{-9 T_0
   \left(17 T_0+32\right) \zeta _0+2 T_0 \big[T_0 \left(9 \Omega _{\Lambda }+33\right)+86\big]+6 \alpha  \big[612 \zeta _0^2-6 \left(24
   \Omega _{\Lambda }+73\right) \zeta _0+ \nonumber \\ & \quad {} +5 \big(14 \Omega _{\Lambda }+T_0 \left(\Omega _{\Lambda }+5\right)+2\big)\big]\big\} \Omega _M+8
   \zeta _0\big\{-15 T_0^2+2 \left(153 \zeta _0-50\right) T_0+ \nonumber \\ & \quad {} +12 [6 (5 \alpha +2) \zeta _0-5]\big\}\Big) K_0^6+18 \Big(324
   \alpha ^2 \Omega _{\Lambda } \left(17 T_0+102 \alpha  \Omega _{\Lambda }+40\right) \Omega _M^4+ \nonumber \\ & \quad {} +6 \big\{3 \left(51 \alpha  \Omega _{\Lambda
   }+2\right) T_0^2+360 \alpha  \Omega _{\Lambda } T_0-360 \beta  \Omega _{\Lambda }+16 \alpha [-18 \zeta _0+9 \alpha  \Omega _{\Lambda }
   \left(-51 \zeta _0+6 \Omega _{\Lambda }+ \right. \nonumber \\ & \quad {} \left. +17\right)+13]\big\} \Omega _M^3+\big\{\big[-153 \zeta _0+18 (5 \alpha +4) \Omega _{\Lambda
   }+107\big] T_0^2+12 \big[50 \alpha  \Omega _{\Lambda }-6 \zeta _0 \left(51 \alpha  \Omega _{\Lambda }+ \right. \nonumber \\ & \quad {} \left. +10\right)+51\big] T_0+24 \alpha 
   \big[153 \zeta _0^2-4 \big(9 (5 \alpha +6) \Omega _{\Lambda }+46\big) \zeta _0+72 \Omega _{\Lambda }+101\big]\big\} \Omega _M^2+2
   \big\{5 T_0^2+ \nonumber \\& \quad {} +12 \zeta _0 \left(51 \zeta _0-44\right) T_0+8 \zeta _0 \left(-75 \alpha +72 \zeta _0-71\right)+2 [T_0 \left(5 T_0-72
   \zeta _0+50\right)-60 \alpha  \zeta _0] \Omega _{\Lambda }+ \nonumber \\& \quad {} +20\big\} \Omega _M+8 \left(15 T_0-102 \zeta _0+50\right) \zeta _0^2\Big)
   K_0^5-27 \Big(1296 \alpha ^2 \Omega _{\Lambda } \left(51 \alpha  \Omega _{\Lambda }+4\right) \Omega _M^5+ \nonumber \\ & \quad {} +9 \Omega _{\Lambda } \big\{3
   \alpha  \big[17 T_0^2+4 \left(51 \alpha  \Omega _{\Lambda }+20\right) T_0+16 \alpha  \big(-51 \zeta _0+15 (\alpha +2) \Omega _{\Lambda
   }+29\big)\big]+ \nonumber \\& \quad {} -160 \beta\big\} \Omega _M^4+4 \big\{9 \Omega _{\Lambda } T_0^2+[-72 \zeta _0+9 \alpha  \Omega _{\Lambda }
   \left(-204 \zeta _0+24 \Omega _{\Lambda }+73\right)+82] T_0+ \nonumber \\ & \quad {} +6 \left[\alpha  \left(180 \alpha -360 \zeta _0+149\right)-60 \beta 
   \left(\Omega _{\Lambda }-1\right)\right] \Omega _{\Lambda }\big\} \Omega _M^3+4 \big\{5 \Omega _{\Lambda } T_0^2+\big[\zeta _0 \left(153
   \zeta _0-214\right)+ \nonumber \\& \quad {} -2 \big(18 (5 \alpha +4) \zeta _0-71\big) \Omega _{\Lambda }+71\big] T_0+360 \zeta _0^2-572 \zeta _0+12 \alpha 
   \big[\zeta _0 \left(153 \zeta _0-50\right)-10 \Omega _{\Lambda }+ \nonumber \\& \quad {} +25\big] \Omega _{\Lambda }+172\big\} \Omega _M^2-16 \zeta _0 \big\{102
   \zeta _0^2-12 \left(3 \Omega _{\Lambda }+11\right) \zeta _0+5 [T_0+2 \left(T_0+4\right) \Omega _{\Lambda }+  \nonumber \\& \quad {}  +2]\big\} \Omega
   _M-160 \zeta _0^3\Big) K_0^4+324 \Omega _M \Big(4131 \alpha ^3 \Omega _{\Lambda }^2 \Omega _M^5+27 \alpha  \Omega _{\Lambda } \big\{72
   \alpha  \Omega _{\Lambda }+T_0 \left(51 \alpha  \Omega _{\Lambda }+ \right. \nonumber \\& \quad \left. +4\right)\big\}\Omega _M^4+3 \Omega _{\Lambda } \big\{\alpha  \big[-360
   \zeta _0+9 \alpha  \left(5 T_0-102 \zeta _0+20\right) \Omega _{\Lambda }+T_0 \left(-153 \zeta _0+72 \Omega _{\Lambda
   }+92\right)+ \nonumber \\& \quad {} +186\big]-90 \beta  \Omega _{\Lambda }\big\} \Omega _M^3+\big\{36 \left(51 \alpha  \Omega _{\Lambda }+2\right) \zeta _0^2-2
   \big[216 \alpha  \Omega _{\Lambda }^2+9 \left(73 \alpha +2 T_0\right) \Omega _{\Lambda }+82\big] \zeta _0+ \nonumber \\& \quad {} +15 \alpha  \left(T_0+14\right)
   \Omega _{\Lambda }^2+\big[90 \alpha +(75 \alpha +61) T_0\big] \Omega _{\Lambda }+112\big\} \Omega _M^2+\big\{5 T_0 \Omega _{\Lambda
   }^2+2 \big[T_0 \left(5+ \right. \nonumber \\ & \quad {} \left. -10 \zeta _0\right)+2 \zeta _0 \big(9 (5 \alpha +4) \zeta _0-61\big)+30\big] \Omega _{\Lambda }-2 \big[\zeta _0
   \big(\zeta _0 \left(51 \zeta _0-107\right)+71\big)+5\big]\big\} \Omega _M+ \nonumber \\& \quad {} +20 \zeta _0^2 \left(2 \Omega _{\Lambda }+1\right)\Big)
   K_0^3-243 \Omega _M^2 \Omega _{\Lambda } \Big(81 \alpha ^2 \Omega _{\Lambda } \left(17 T_0+68 \alpha  \Omega _{\Lambda }+40\right) \Omega
   _M^4+ \nonumber \\& \quad {} +12 \big\{2 \alpha  \big[-36 \zeta _0+9 \Omega _{\Lambda } \big(T_0+\alpha  \left(-51 \zeta _0+6 \Omega _{\Lambda
   }+17\right)\big)+26\big]-45 \beta  \Omega _{\Lambda }\big\} \Omega _M^3+ \nonumber \\ & \quad {} +4 \big\{5 T_0 \left(3 \alpha  \Omega _{\Lambda }+1\right)+3
   \alpha  \big[153 \zeta _0^2-2 \big(9 (5 \alpha +8) \Omega _{\Lambda }+92\big) \zeta _0+47 \Omega _{\Lambda }+101\big]\big\} \Omega
   _M^2+ \nonumber \\ & \quad {} +4 \big\{4 \zeta _0 \left(9 \zeta _0-23\right)-30 \alpha  \zeta _0 \left(\Omega _{\Lambda }+5\right)+5 T_0 \left(3 \Omega _{\Lambda
   }-1\right)+56\big\} \Omega _M-40 \zeta _0 \left(-2 \zeta _0+\Omega _{\Lambda }+ \right. \nonumber \\ & \quad {} \left.  +2\right)\Big) K_0^2+1458 \Omega _M^3 \Omega _{\Lambda }^2
   \Big(-40 \zeta _0+\Omega _M \big\{10 T_0-90 \beta  \Omega _M \left(\Omega _M+\Omega _{\Lambda }-1\right)+3 \alpha  \big[-20 \zeta _0+ \nonumber\\ & \quad {} +3
   \Omega _M \big(3 \alpha  \left(\Omega _M+2\right) \left(12 \Omega _M+5\right)-3 \zeta _0 \left(51 \alpha  \Omega _M+8\right)+4\big)+2
   \big(9 \alpha  \Omega _M^2 \left(51 \Omega _M \alpha +5 \alpha + \right. \nonumber \\ & \quad {} \left. +12\right)-10\big) \Omega _{\Lambda }+50\big]\big\}+10\Big) K_0-2187
   \Omega _M^4 \Omega _{\Lambda }^2 \Big(-20 \zeta _0 + \nonumber \\ & \quad {}+3 \Omega _M \big\{\alpha  \big[\Omega _M \big(3 \alpha  \left[9 \Omega _M \left(17
   \alpha  \Omega _M+4\right)+10\right] \Omega _{\Lambda }-20\big)+20\big]-30 \beta  \Omega _M \Omega _{\Lambda }\big\}+ \nonumber \\ & \quad {} +20\Big)\bigg]
   \overline{H}_0^4 -8 \overline{\Omega}_M H_0^2 \Omega _M \mathcal{A} \bigg[576 \alpha ^2 \delta _1 K_0^8-32 \alpha  \Big(\delta _1 \left(54 \Omega
   _M \alpha -20 \alpha +9 T_0+8\right)+ \nonumber \\ & \quad {} -8 \alpha  \delta _2\Big) K_0^7+4 \Big(\delta _1 \big\{4 \Omega _M \left(81 \Omega _M-80\right)
   \alpha ^2+8 \big[3 \left(9 T_0+16\right) \Omega _M-2 \left(5 T_0+27 \zeta _0+ \right. \nonumber \\ & \quad {} \left. +5\right)\big] \alpha +T_0 \left(9 T_0+16\right)\big\}-32
   \alpha  \delta _2 \left(T_0+3 \alpha  \Omega _M+1\right)\Big) K_0^6+4 \Big(4 \delta _2 \big\{T_0 \left(T_0+2\right)+ \nonumber \\ & \quad {}+12 \alpha 
   \big[\left(T_0+3\right) \Omega _M-4 \zeta _0\big]\big\}+\delta _1 \big\{\left(10-27 \Omega _M\right) T_0^2+2 \big[54 \zeta _0+\Omega _M
   \left(-81 \Omega _M \alpha +80 \alpha + \right. \nonumber \\ & \quad {} \left. -48\right)+10\big] T_0+48 \zeta _0 \left(27 \Omega _M \alpha -10 \alpha +2\right)-8 \alpha  \Omega
   _M \big[90 \Omega _M+9 \left(27 \alpha  \Omega _M+4\right) \Omega _{\Lambda }+ \nonumber \\ & \quad {} -40\big]\big\}\Big) K_0^5+\Big(\delta _1 \big\{864
   \alpha  \left(27 \alpha  \Omega _{\Lambda }+2\right) \Omega _M^3+3 \big[27 T_0^2+24 \left(27 \alpha  \Omega _{\Lambda }+10\right) T_0+ \nonumber \\& \quad {} -16
   \alpha  \big(81 \zeta _0+36 (5 \alpha -3) \Omega _{\Lambda }+25\big)\big]\Omega _M^2-16 \big[5 T_0^2+\left(81 \zeta _0-18 \Omega
   _{\Lambda }+20\right) T_0+144 \zeta _0+ \nonumber \\ & \quad {} +30 \alpha  \left(-8 \zeta _0+5 \Omega _{\Lambda }+1\right)-1\big] \Omega _M+48 \zeta _0 \left(10
   T_0+27 \zeta _0+10\right)\big\}-24 \delta _2 \big\{\Omega _M \big[T_0^2+6 T_0+ \nonumber \\ & \quad {} +24 \alpha  \left(6 \alpha  \Omega _{\Lambda } \Omega
   _M+\Omega _M+\Omega _{\Lambda }+1\right)\big]-8 \zeta _0 \left(T_0+6 \alpha  \Omega _M+1\right)\big\}\Big) K_0^4+ \nonumber \\ & \quad {} -12 \Big(1458 \alpha
   ^2 \delta _1 \Omega _{\Lambda } \Omega _M^4+18 \big\{T_0 \delta _1 \left(27 \alpha  \Omega _{\Lambda }+2\right)-4 \alpha  \big[5 (4 \alpha
   -3) \delta _1+6 \alpha  \delta _2\big] \Omega _{\Lambda }\big\} \Omega _M^3+ \nonumber \\ & \quad {} +\big\{-12 \delta _2 \big[T_0+6 \alpha  \left(T_0+2\right)
   \Omega _{\Lambda }\big]-\delta _1 \big[T_0 \big(81 \zeta _0+36 (5 \alpha -2) \Omega _{\Lambda }+25\big)+4 \big(120 \alpha  \Omega
   _{\Lambda }+ \nonumber \\ & \quad {} +9 \zeta _0 \left(27 \alpha  \Omega _{\Lambda }+10\right)-4\big)\big]\big\} \Omega _M^2-12 \delta _2 \big\{-6 \zeta _0+T_0
   \left(-2 \zeta _0+\Omega _{\Lambda }+1\right)+2\big\} \Omega _M+ \nonumber \\ & \quad {} -2 \delta _1 \big\{-162 \zeta _0^2+8 \big[9 \Omega _{\Lambda }-5
   \left(T_0+2\right)\big] \zeta _0+5 \big[T_0+\left(5 T_0+4\right) \Omega _{\Lambda }+2\big]\big\} \Omega _M-24 \left(5 \delta _1+ \right. \nonumber \\ & \quad {} \left. +2
   \delta _2\right) \zeta _0^2\Big) K_0^3+18 \Omega _M \Big(\delta _1 \big\{27 \alpha  \Omega _{\Lambda } \left(9 T_0+54 \alpha  \Omega
   _{\Lambda }+40\right) \Omega _M^3+12 \big[-12 \zeta _0+3 T_0 \Omega _{\Lambda }+ \nonumber \\ & \quad {} +2 \alpha  \Omega _{\Lambda } \big(-10
   \left(T_0+2\right)-81 \zeta _0+18 \Omega _{\Lambda }\big)+2\big] \Omega _M^2-2 \big[-\zeta _0 \left(81 \zeta _0+50\right)+\big(25
   T_0+ \nonumber \\ & \quad {} +24 \left[(6-15 \alpha ) \zeta _0+2\right]\big) \Omega _{\Lambda }+11\big] \Omega _M+40 \zeta _0 \left(-4 \zeta _0+5 \Omega _{\Lambda
   }+1\right)\big\}-12 \delta _2 \big\{6 \alpha  \left(T_0+ \right. \nonumber \\ & \quad {} \left. +3\right) \Omega _{\Lambda } \Omega _M^2+\big[T_0 \Omega _{\Lambda }-4 \zeta _0
   \left(6 \alpha  \Omega _{\Lambda }+1\right)+4\big] \Omega _M+4 \zeta _0 \left(\zeta _0-\Omega _{\Lambda }-1\right)\big\}\Big) K_0^2+ \nonumber \\ & \quad {} +108
   \Omega _M^2 \Omega _{\Lambda } \Big(\delta _1 \big\{36 \Omega _M \zeta _0-50 \zeta _0+4 \Omega _M+3 \alpha  \Omega _M \big[-80 \zeta
   _0+\left(81 \zeta _0-36 \Omega _M+25\right) \Omega _M+ \nonumber \\ & \quad {} +10\big]+\big[3 \alpha  \Omega _M \big(50-9 \Omega _M \left(27 \Omega _M \alpha -10
   \alpha +8\right)\big)+20\big] \Omega _{\Lambda }+10\big\}-12 \delta _2 \big\{6 \alpha  \Omega _M \zeta _0+ \nonumber \\ & \quad {} +\zeta _0-3 \alpha  \Omega _M
   \big[\Omega _{\Lambda }+\Omega _M \left(3 \alpha  \Omega _{\Lambda }+1\right)+1\big]-1\big\}\Big) K_0+81 \Omega _M^3 \Big(\delta _1
   \big\{3 \alpha  \Omega _M \big[3 \alpha  \left(81 \Omega _M+ \right. \nonumber \\ & \quad {} \left. -80\right) \Omega _M+72 \Omega _M-100\big]-20\big\}-72 \alpha  \delta _2
   \Omega _M \left(3 \alpha  \Omega _M+1\right)\Big) \Omega _{\Lambda }^2\bigg] \overline{H}_0^2+ \nonumber \\& \quad {} -80 H_0^4 \Omega _M^2 \bigg[9 \Omega _M^2-2
   \left(6 K_0+5\right) \Omega _M+4 K_0 \left(K_0+1\right)\bigg] \mathcal{A}{}^3 \Bigg\} \,,   
\end{align}
where 
\begin{align}
\mathcal{A} &= -4 \alpha  K_0^3+6 K_0 \left(\zeta _0-\Omega _M\right)+K_0^2 \left(T_0+2\right)+9
   \Omega _{\Lambda } \Omega _M \left(3 \alpha  \Omega _M+1\right)\, .
\end{align}

\acknowledgments  

\bibliographystyle{h-physrev4}
\bibliography{mybib}
\end{document}